\documentclass[aps,pra,twocolumn,showpacs,superscriptaddress]{revtex4-1}

\usepackage{amsfonts,mathrsfs,amsmath,amsthm,amssymb,bbold}
\usepackage{epsfig}
\usepackage{bm}
\usepackage{tabulary}
\usepackage{color}
\usepackage{xcolor,verbatim}
\graphicspath{{./images/}}
\DeclareMathOperator{\Tr}{Tr}

\begin{document}  
\title {\bf Direct state measurements under state-preparation-and-measurement errors} 

\author{Kieu Quang Tuan}
\affiliation{University of Science, VNUHCM, Ho Chi Minh City, Vietnam}
\author{Hung Q. Nguyen}
\affiliation{Nano and Energy Center, VNU University of Science, Vietnam National University, 120401, Hanoi, Vietnam}

\author{Le Bin Ho}
\affiliation{Ho Chi Minh City Institute of Physics, VAST, Ho Chi Minh City, Vietnam}
\affiliation{Research Institute of Electrical Communication, 
Tohoku University, Sendai, 980-8577, Japan}
\thanks{Electronic address: binho@riec.tohoku.ac.jp}

\date{\today}

\begin{abstract}
Direct state measurement (DSM)
 is a tomography method that allows 
 for retrieving quantum states' wave functions directly.
However, a shortcoming of current studies on 
the DSM is that 
it does not provide access 
to noisy quantum systems. 
Here, we attempt to fill the gap 
by investigating the DSM measurement precision 
that undergoes the state-preparation-and-measurement (SPAM) errors.
We manipulate a quantum controlled measurement framework
with various configurations and 
compare the efficiency between them. 
Under such SPAM errors,
the state to be measured lightly deviates from the true state, 
and the measurement error in the postselection process 
results in less accurate in the tomography. 
Our study could provide a reliable tool for 
SPAM errors tomography and contribute to understanding 
and resolving an urgent demand for 
current quantum technologies.
\end{abstract}
%
%

\maketitle

\section {Introduction}\label{sec_i}
\sloppy
In quantum measurement,
a process for retrieving the wave functions of 
quantum states is known as 
quantum state tomography (QST)~\cite{Paris2004}. 
It plays a crucial role in a wide range 
of quantum technologies, 
from randomized benchmarking~\cite{Helsen2019}, 
calibrating quantum operations~\cite{Frank2017}, 
to experimentally validating 
quantum computing devices~\cite{Gheorghiu2019}.

However, the most severe obstacle against the
realizing quantum devices, such as
the current version of 
noisy intermediate-scale quantum (NISQ) 
computers,
is that the noise can 
result in the superposition 
and entanglement loss. 
Typically, the noise can occur 
by the imperfection 
over controlling the system
\cite{Preskill2018quantumcomputingin}
or 
when the system inevitably interacts 
with its surrounding environment
\cite{schlosshauer_2007}.
Such noises appear during the 
preparation and measurement processes
can be referred to as 
state-preparation-and-measurement 
(SPAM) errors~\cite{PhysRevA.92.042312}
and thus limit the accuracy of quantum tomography
(or quantum measurement in general.) 
In this work, we model and evaluate the effect of
such SPAM errors in 
a direct state measurement scheme.

Besides the conventional quantum state tomography
\cite{Kosaka2009,Vanner2013,PhysRevLett.116.230501,
PhysRevA.93.052105}
in which particularly challenging for large systems,
a direct state measurement (DSM) method was proposed 
\cite{Lundeen2011,PhysRevLett.108.070402}
and extensively used.
This method is straightforward, versatile, simple,
and has been numerously applied in large systems  
\cite{Shi:15,PhysRevLett.113.090402,
Malik2014,Bolduc2016,PhysRevA.98.023854},
mixed quantum states  
\cite{PhysRevLett.108.070402,PhysRevLett.117.120401,
PhysRevLett.121.230501},
enlarged Hilbert space \cite{HO2019289},
and nonlocal entangled states 
\cite{PhysRevLett.123.150402}.
Further study on the novelty, efficacy,
and significance of the DSM had been reported 
\cite{PhysRevA.92.062133}.
So far, the statistical error 
and systematic error
have been examined
\cite{Sainz_2016,PhysRevA.94.012329,Ho_2020}.

The measurement employed in 
the DSM follows the von Neumann interaction 
and requires a postselection technique~\cite{vonNeumann}. 
Recently, however, a quantum controlled measurement framework
has been proposed \cite{F_Hofmann_2014, Ogawa_2019} 
and can be used for analyzing systematic errors in the DSM
\cite{Ho_2020}. This framework contains 
a probe-controlled-system type of the interaction 
where a qubit probe controls 
a target system through the evolution~\cite{Ogawa_2019}
\begin{align}\label{eq:inter}
\bm{U} = \bm U_0\otimes|0\rangle\langle 0|
+ \bm U_1 \otimes|1\rangle\langle 1|,
\end{align}
where $\bm U_0, \bm U_1$ 
are the operators of the target system, 
and $|0\rangle, |1\rangle$ 
are two elements of the basis in the control qubit probe.
Such an interaction can be realized 
by using a Fredkin gate,
as illustrated in Fig.~\ref{fig1}(a).
This measurement scheme is operationally equivalent
to  the von Neumann interaction \cite{Ho_2020}.
Moreover, when combining with 
the postselection technique, 
it allows for cyclic transforming 
the system operators, 
which results in different configurations
\cite{Ogawa_2019}.
See, for example, configurations C1, 
C2 in Fig.~\ref{fig1}(c, d) below. 
In some configurations,
a scan-free technique \cite{Shi:15}
can be used directly, which assists in improving 
the measurement 
precision.

In this work, we numerically 
and analytically investigate 
the measurement efficiency in the DSM
that both the pre- and postselected states 
are under the noises during
the preparation and measurement (postselection) processes. 
These noises can be seen as SPAM errors.
We consider two cases of noise: 
(i) noisy in pure states due to the imperfection, and 
(ii) noisy in mixed states due to the interaction 
with the surrounding environment. 
Previously, Shikano and Hosoya had proposed 
such a noisy system due to its interaction 
with the surrounding
environment 
\cite{Shikano_2009}.

We employ the quantum controlled 
measurement framework with 
different configurations under 
SPAM errors and compare their efficiencies. 
We use the mean trace distance 
obtained from the Monte Carlo simulation to 
evaluate the  measurement precession. 
Noting that through this paper,
we refer to the SPAM errors as ``noises"
that occur during the state-preparation
and postselection processes. 
Besides, we also use the 
``systematic error" and ``statistical error"
terminologies for the measurement precision.
The systematic error is
typically caused by
the different configurations 
of the measurement scheme~\cite{Ho_2020},
while the statistical error regards the 
finite number of the repeated measurements.

The rest of this paper is organized as follows. 
In Sec.~\ref{secii}, we define SPAM errors 
and two configurations of the 
quantum controlled measurement framework for
pure states 
and then extend to mixed states.
The numerical results are presented in Sec.~\ref{seciii}
for both cases 
of pure and mixed quantum states. 
In Sec.~\ref{seciv}, we analytically investigate 
the impact of the state-preparation error on
the precision in the reconstruction process. 
Finally, we summarize our work in Sec.~\ref{secv}. 
 
{\section{Direct state measurement
with quantum controlled interaction}\label{secii}}

We first consider the SPAM errors 
for pure-state quantum systems 
and then generalize to mixed states.
Typically, a direct state measurement framework includes 
(i) the coupling between a quantum target system
and a control qubit probe, 
(ii) the postselection of the target 
system onto a final state, 
(iii) and the measurement on the control qubit probe.

\subsection{State-preparation and 
state-postselection errors for pure states}\label{sec_iiA}

Assume that a pure quantum state 
is spanned in a computational basis 
$\{|n\rangle\}_0^{d-1}$ 
of a system $d$-dimensional space as follows 
\begin{align}\label{eq:psi}
|\psi\rangle = \sum_{n=0}^{d-1}\psi_n|n\rangle,
\end{align}
where $\psi_n \equiv \langle n|\psi\rangle$ 
is a complex amplitude and satisfies 
$\sum_n|\psi_n|^2 = 1$. 
We model a noise
in the state-preparation system 
due to the ``imperfection" 
during the preparation process
\cite{Preskill2018quantumcomputingin}.
See an example in App.~\ref{appA}.
In general, we assume that
under such an imperfection,
the original quantum state transforms into
\begin{align}\label{e:pp}
|\psi\rangle\to|\psi'\rangle
= \dfrac{1}{\mathcal N}\sum_n
\bigl(\psi_n + \delta_n\bigr)|n\rangle,
\end{align}
where $\mathcal N$ is a normalization  constant,
and $\delta_n$ 
is a complex random perturbation,
i.e., $\delta_n = x_1 + ix_2$,
where $x_1, x_2$ are random numbers 
and obey the normal distribution with mean zero,
such that $f(x)= \frac{1}{\sigma\sqrt{2\pi}}
\exp[-\frac{1}{2}(\frac{x}{\sigma})^2]$.
Here, 
$\sigma$ is the standard derivation, which stands for the 
noise parameter.  
Under such perturbation, 
$|\psi'\rangle$ can be seen as a 
``probability distribution" of state $|\psi\rangle$. 
Here, we consider this simple case
to evaluate the efficiency  
of the proposed protocol
for pure states.

\begin{figure} [t]
\centering
\includegraphics[width=8.6cm]{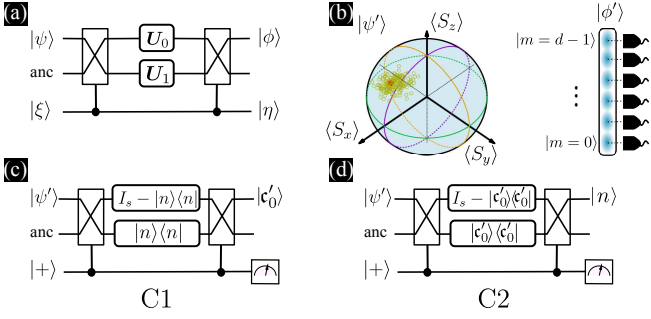}
\caption{
(a) A quantum controlled measurement scheme 
for the direct state measurements:
$|\psi\rangle$ and $|\phi\rangle$ are pre-
and postselected states in the target system
which will become $|\psi'\rangle$ and $|\phi'\rangle$
under the SPAM errors;
$|\xi\rangle$ and $|\eta\rangle$ are the initial
and final states in the control qubit probe.
The action of $\bm U_0$ or $\bm U_1$ onto
the target state $|\psi\rangle$ is controlled 
by the control qubit probe via a Fredkin gate
with an ancillary (anc) system.
(b) A visualization of a noisy
state-preparation $|\psi'\rangle$. 
The red-filled circle is the original state, and 
the open circles stand for the noisy states, 
which lightly deviate from the original one.
Here $\langle S_i\rangle, i = \{x, y, z\}$ 
are the expectation 
values of the total spin operators.
The noisy state-postselection 
$|\phi'\rangle$ is also
illustrated.
(c, d) Two configurations of 
the quantum controlled measurement scheme. 
In (c), when the control qubit probe
is in state $|0\rangle$, 
the operator $\bm U_0 = \bm {I}_s - |n\rangle\langle n|$
will operate on the target system;
while that will be $\bm U_1 = |n\rangle\langle n|$ 
when the state is $|1\rangle$
(the target state will be swapped). The target system afterward 
is postselected onto the final state $|\mathfrak{c}_0'\rangle$.
The role of $|n\rangle$ and $|\mathfrak{c}_0'\rangle$ 
can be interchanged to form the configuration C2 in (d).
}
\label{fig1}
\end{figure}

A spin-based visualization for $|\psi\rangle$
and an ensemble of $|\psi'\rangle$'s
is illustrated in Fig. \ref{fig1}(b). 
Therein, the red-filled circle is the true state
$|\psi\rangle$, and the open circles are 
the statistical distribution of $|\psi'\rangle$'s.
We emphasize that the noise parameter is unknown,
therein
it lightly deviates the quantum state $|\psi'\rangle$ from
the original quantum state $|\psi\rangle$. 
Therefore, to evaluate the measurement efficiency 
under the noise, we need to compare the reconstructed state 
with the true state.
This method has widely used in 
the QST under noise
\cite{PhysRevResearch.1.033157,Palmieri2020}.

For the noisy state-postselection, 
in the same spirit as the noisy state-preparation, 
we assume that the postselected state $|\phi\rangle$
contains a small noise,
and thus it becomes 
\begin{align}\label{eq:post'}
|\phi'\rangle\equiv |\mathfrak{c}'_0\rangle 
= \frac{1}{\mathcal{M}}\sum_{m}
(1+\kappa_m)|m\rangle
= \sum_{m} \mathfrak{c}_m |m\rangle,
\end{align}
where $\mathcal{M}$ is 
the normalization factor, 
$\mathfrak{c}_m = \frac{1+\kappa_m}{\mathcal{M}}$,
and $\kappa_m$ is a random number.
We illustrate the noisy state-postselection 
in Fig.~\ref{fig1}(b). If a detector is perfect, 
it will detect exactly the state (or position)
at each $|m\rangle$. Inversely,
if there is imperfection, the detector 
will detect a biased state,
i.e., $|m\rangle + \kappa_m|m\rangle$.
Without loss of generality, we can assume $\kappa_m$ 
is real and follows the
normal distribution because the complex
part of the postselected state 
can be absorbed into the phase  
(see Eq.~\eqref{eq:kk}.)
Together, the noisy state-preparation and 
state-postselection form SPAM errors.

This simple case of the SPAM errors is
used for evaluating the performance 
of the measured protocol
for pure states.
Besides, such imperfections 
are widespread 
in the current version of 
the NISQ
computers~\cite{Preskill2018quantumcomputingin}.
It is thus helpful to study such noise
in quantum tomography, an essential aspect 
of the NISQ.

Furthermore, similar to Ref.~\cite{Shikano_2009}, 
here we draw our attention to the role of the SPAM errors 
in the target system, 
which affects the tomography process itself. 
We thus, restrict ourselves to the case 
where the control qubit 
is pure and perfectly prepared. 
In practical realization, 
the qubit probe can be prepared 
such that it is error-free. 
For example, in photonics systems, 
the control qubit probe is the photon polarization mode, 
which can be realized using polarizers (optical filters)
\cite{Ogawa_2019,Shi:15}. 
Hence, the noise in such a qubit probe can be eliminated. 

%

\subsection{Quantum controlled measurements 
for pure states}
We first consider Configuration 1 (or C1 for short),
described by 
an interaction scheme 
between a target system 
and a control qubit probe.
The measurement scheme is 
schematically shown in Fig.~\ref{fig1}(c). 
The initial state of the target system 
is $|\psi'\rangle$, while 
the control qubit is initially 
prepared in the state $|\xi\rangle$,
i.e., $|\xi\rangle \equiv |+\rangle = 
\bigl(|0\rangle + |1\rangle\bigr)/\sqrt{2}$.
The initial joint state becomes
%
$|\Psi\rangle 
 = \sum_n \psi_n'|n\rangle \otimes |+\rangle,$
%
where $\psi'_n = (\psi_n + \delta_n)/\mathcal{N}$. 
The interaction between the target system and
the control qubit probe 
is given by
\begin{align}\label{eq:U-io}
U_n = 
\bigl(\bm{I}_{\rm s}-|n\rangle\langle n|\bigr)
\otimes |0\rangle\langle 0| + |n\rangle\langle n|
\otimes |1\rangle\langle 1|,
\end{align}
where $\bm{I}_{\rm s}$ is the identity matrix
in the target system. 
The action of Eq.~\eqref{eq:U-io}
on the probe state $|\xi\rangle$ 
 is operationally equivalent to 
a von Neumann measurement
given by a strong interaction 
$U = \exp(-i\frac{\pi}{2} 
|n\rangle\langle n|\otimes\sigma_y)
$
acts on the 
control qubit prepared in $|0\rangle$
\cite{Ogawa_2019,Ho_2020}.
Particularly, we have
$ U(\bm I_{\rm s}\otimes |0\rangle) = 
(\bm I_{\rm s} - |n\rangle\langle n|)\otimes|0\rangle
+ |n\rangle\langle n|\otimes
|1\rangle$.
We thus can choose $U_0 = \bm I_{\rm s} - |n\rangle\langle n|$
and $U_1 = |n\rangle\langle n|$ 
as can be seen in Eq.~\eqref{eq:U-io}.

%

After the interaction, 
%
we postselect the target system onto the conjugate basis
$|\mathfrak{c}'_0\rangle$ as given in Eq.~\eqref{eq:post'}.
Then, the final control qubit state is given by
(see App.~\ref{appB})
\begin{align}\label{eq:final}
|\eta\rangle=\dfrac{1}{\sqrt{2}}\Bigl[
\big(\Gamma - \mathfrak{c}_n\psi'_n\big)|0\rangle
+\mathfrak{c}_n\psi'_n|1\rangle\Bigr],
\end{align} 
where we have set 
$\Gamma = \sum_{m}\mathfrak{c}_{m}\psi'_{m}$,
which can be chosen to be real 
\cite{PhysRevLett.116.040502}.

Finally, we measure the control qubit probe 
in the Pauli basis $\{|j\rangle\}$,
where $|j\rangle \in \{ |0\rangle, |1\rangle, 
|+\rangle, |-\rangle, |L\rangle, |R\rangle\}$,
$|\pm\rangle = \frac{1}{\sqrt{2}}\bigl(|0\rangle\pm|1\rangle\bigr),
|L\rangle = \frac{1}{\sqrt{2}}\bigl(|0\rangle + i|1\rangle\bigr), 
|R\rangle = \frac{1}{\sqrt{2}}\bigl(|0\rangle - i|1\rangle\bigr)$.
The corresponding probability is
$P_j = \langle \eta|j\rangle\langle j|\eta\rangle = 
|\langle j|\eta\rangle|^2$,
where the subscript $j$ takes 0, 1, +, -, $L$, and $R$
corresponding to the elements in the Pauli basis.
Then, the real and imaginary parts of the amplitude
$\psi'_n$ are reproduced as
\begin{align}\label{eq:rec}
{\rm Re}\psi'_n=\frac{P_+-P_-+2P_1}
{\mathfrak{c}_n\Gamma}, \
{\rm Im}\psi'_n=\frac{P_L-P_R}
{\mathfrak{c}_n\Gamma}.
\end{align}

In this scheme, it is worth noting that 
after postselecting the target system onto 
$|\mathfrak{c}'_0\rangle$, we discard all these other 
results and then repeat
the measurement for all $\{n\}$.

Next, we describe Configuration 2 (or C2 for short). 
In this case, we interchange the role of $|n\rangle$
and $|\mathfrak{c}'_0\rangle$ as shown in Fig. \ref{fig1}(d).
The interaction is given by
\begin{align}\label{eq:U-i}
U=(\bm{I}_{\rm s}-|\mathfrak{c}'_0\rangle
\langle \mathfrak{c}'_0|)\otimes|0\rangle\langle0|
+|\mathfrak{c}'_0\rangle
\langle \mathfrak{c}'_0|\otimes|1\rangle\langle1|.
\end{align}
The random noise $\kappa_m$ that appear
in $|\mathfrak{c}'_0\rangle$ is unknown.
Therefore, in practice, this interaction can be done by
applying the state $|\mathfrak{c}_0\rangle$ and 
$|\mathfrak{c}_0\rangle^\perp 
= \bm{I}_{\rm s}-|\mathfrak{c}_0\rangle
\langle \mathfrak{c}_0|$ onto the
target system while $\kappa_m$ is ``self-rising"
during the measurement process.

After the interaction, the target system is postselected 
onto the basis $|n\rangle$ while the remaining state of the control qubit probe
is given by
\begin{align}\label{eq:f-II}
    |\eta\rangle=\dfrac{1}{\sqrt{2}}\Bigl[\big(
    \psi'_n-\mathfrak{c}_n\Gamma\big)|0\rangle+
\mathfrak{c}_n\Gamma|1\rangle\Big].
\end{align}
Measuring the control qubit probe in the Pauli basis
as above, we obtain
\begin{align}\label{eq:recc2}
{\rm Re}\psi'_n=\frac{P_+-P_-+2P_1}
{\mathfrak{c}_n\Gamma},\
{\rm Im}\psi'_n=\frac{P_L-P_R}
{\mathfrak{c}_n\Gamma}.
\end{align}
(See App.~\ref{appB} for detailed calculation.)  

We emphasize that different from C1, here,
we keep all the postselected state $\{|n\rangle\}$.
This technique is known as ``scan-free,"
and can be employed by using 
an array 
detector~\cite{Shi:15,Ogawa_2019}.

\subsection{Quantum controlled 
measurements for mixed states}
We now consider the general case of 
SPAM errors for mixed states.
Assume the initial target system state
is prepared in $\rho_0$. It will transform into
$\rho'_0$ by passing through the noise channel 
$\mathcal{E}$, 
such that~\cite{nielsen_chuang_2000}
\begin{align}\label{eq:rhot}
\rho'_0 \equiv \mathcal{E}(\rho_0) = 
\sum_k E_k\rho_0 E_k^\dagger,
\end{align}
where $E_k$ is an operation element
satisfying the completeness
relation $\sum_k E_k^\dagger E_k = \bm I$.
Noting that this noise model
can be applied for any noise.

Then, the state of the joint target-control system is
\begin{align}
\Lambda = \rho'_0 \otimes |+\rangle\langle +|, 
\end{align}
where
$\rho'_0$ 
can be expressed as $\sum_{n,m=0}^{d-1}
\rho'_{nm}|n\rangle\langle m|$, which
is the state to be reconstructed after passing through 
the noisy channel, 
and $|+\rangle$ is the initial state of the control qubit probe.
Here, we will reconstruct the components $\rho'_{nm}$.

For C1, the target-control interaction 
is the same as pure-state case, i.e.,
 $
U_n = \big(\bm{I}_{\rm s} - |n\rangle\langle n|\big) 
\otimes |0\rangle\langle 0 |+ 
|n\rangle\langle n|\otimes |1\rangle\langle 1| 
$,
and the postselected state under the
noise is given by
\begin{align}\label{eq:kk}
|\mathfrak{c}'_k\rangle\langle \mathfrak{c}'_k| = 
\sum_{n,m} e^{i2\pi (m-n)k/d}
\mathfrak{c}_m
\mathfrak{c}_n
|m\rangle\langle n|, 
\end{align}
where $\mathfrak{c}_m = 
\frac{1+\kappa_{m}}{\mathcal{M}}$.
Using Fourier transformation, 
we obtain the reconstructed state as
\begin{align}
    \rho'_{nm} \varpropto \frac{1}
    {\mathfrak{c}_{n}\mathfrak{c}_{m}}
    \Big[d\delta_{n,m}
    \Lambda''_{11}(n,k)+\sum_{k=0}^{d-1}e^{\frac{i2\pi (n-m)k}{d}}
    \Lambda_{10}''(n,k) \Big],
\end{align}
where $\Lambda''_{ij}(n,k)$ is a component 
in the final probe state. 
(See App.~\ref{appC} for detailed calculation.)

For C2, the interaction is 
$ 
U_k = \big({\bm I}_{\rm s} - 
|\mathfrak{c}'_k\rangle\langle \mathfrak{c}'_k|\big) \otimes 
|0\rangle\langle 0 |+ |\mathfrak{c}'_k\rangle
\langle \mathfrak{c}'_k|\otimes
|1\rangle\langle 1|,
$
and the postselected state is 
$|n\rangle\langle n|$.
The reconstructed state is given by
\begin{align}
&\rho'_{nm} \varpropto \frac{1}
{\mathfrak{c}_{n}\mathfrak{c}_{m}}
\Big[\sum_{k=0}^{d-1} e^{\frac{i2\pi k(n-m)}{d}}
\Big(\Lambda''_{01}(n,k) + \Lambda''_{11}(n,k) \Big)
\Big].
\end{align}
(See App.~\ref{appC} for detailed calculation.)
The reconstructed state $\rho'$ whose elements
are $\rho'_{nm}$ is not a physical state. 
The normalized state of 
$\rho'$ is denoted as $\widetilde\rho$
and given by
$\widetilde\rho = \rho'^\dagger\rho'/ 
{\rm Tr} (\rho'^\dagger\rho')$, 
which is a normalized Hermitian matrix,
and thus it is a physical state~\cite{Paris2004}.

We emphasize that in the mixed-state case,
both C1 and C2 can implement the scan-free technique
because we can keep all postselected states 
$\{|\mathfrak{c}'_k\rangle\langle\mathfrak{c}'_k|\}$
in C1 and all postselected states $\{|n\rangle\langle n|\}$
in C2 without discarding them.

\section{Numerical results}\label{seciii}
\subsection{Numerical results for pure states}
We employ a simulation scheme using the 
cumulative distribution function (cdf) 
\cite{PhysRevA.89.022122,HO2019289},
that built in to $tqix$ \cite{HO2021107902}. 
To evaluate the accuracy of the reconstruction, 
we use the trace distance as a figure of merit, 
defined by 
\begin{align}\label{trd}
D(\psi,\widetilde\psi')= 
\sqrt{1-|\langle{\widetilde\psi'}|\psi\rangle|^2},
\end{align}
where $|\widetilde{\psi'}\rangle$ is the reconstructed state
from $|\psi'\rangle$,
and $|\psi\rangle$ is the true state 
(see Eq.~\ref{e:pp}.) 
The reconstructed state is calculated from 
Eq.~\eqref{eq:rec} and Eq.~\eqref{eq:recc2} 
for C1 and C2, respectively, 
which provides information about the true
quantum state from the measurement results.
By averaging $D(\psi,\widetilde\psi')$ over 
many repetitions of the measurement procedure,
we obtain the mean trace distance 
$\overline{D}(\psi,\widetilde\psi')$,
which yields slightly different of the true state
\cite{Paris2004}. 

\begin{figure} [t]
\centering
\includegraphics[width=8.6cm]{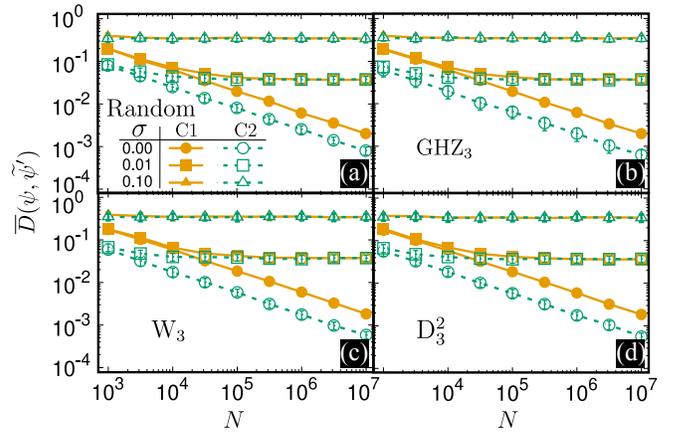}
\caption{Log-log plot of the mean trace distance 
$\overline{D}(\psi,\widetilde\psi')$
as a function 
of the number of copies $N$ for various standard states:
(a) random state,
(b) $\text{GHZ}_{3}$ state,
(c) $\text{W}_{3}$ state,
(d) ${\rm D}_{3}^2$ state.
For each case, the results from C1 (filled symbols)
and C2 (open symbols) are plotted for several errors:
$\sigma = 0.0, 0.01$ and 0.10.
}
\label{fig2}
\end{figure}

First, we numerically examine the mean trace distance
$\overline{D}(\psi,\widetilde\psi')$
versus the number of copies $N$.
The results are shown in Fig.~\ref{fig2} for
several well-known quantum states: (a) random state 
(distributed following Haar measure \cite{9eafeb2573aa4d7a9d3f0f17ec8c9af5}), 
(b) Greenberger-Horne-Zeilinger $\text{GHZ}_3$ state, i.e.,
$ \text{GHZ}_3 = 1/\sqrt{2} (|000\rangle + |111\rangle$),
(c) $\text{W}_3$ state, i.e.,
$ \text{W}_3 = 1/\sqrt{3} (|001\rangle + |010\rangle
+ |100\rangle$),
and (d) Dicke state $D_3^2$, i.e.,
$ \text{D}_3^2 = 1/\sqrt{3} (|011\rangle + |101\rangle
+ |110\rangle$).
For each state, we consider several 
noisy parameters $\sigma = 0.00, 0.01, 0.10$
as examples. 
For simplicity, we choose the same noise parameter 
($\sigma$)
for the state preparation and state postselection. 

For all cases without noise ($\sigma = 0.0$), 
$\overline{D}(\psi,\widetilde\psi')$
continuously decreases as $N$ 
increases, and C2 shows
better results than C1. 
This observation can be understood as a consequence 
of the scan-free process in C2, where all postselected 
states $\{|n\rangle\}$ are kept for the reconstruction.
This is a kind of ``systematic error", 
which typically depends on
the different configurations 
of the measurement scheme \cite{Ho_2020}.
The systematic error cannot be eliminated
even though increasing 
the number of copies $N$ (to eliminate 
the statistical error)
as can be seen from the figure.

\begin{figure} [t]
\centering
\includegraphics[width=8.56cm]{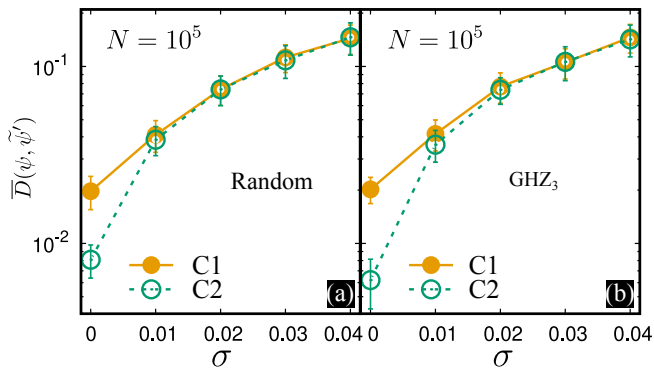}
\caption{Mean trace distance 
$\overline{D}(\psi,\widetilde\psi')$
as a function of 
the noise parameter $\sigma$ for two cases:
(a) random state,
(b) $\text{GHZ}_{3}$ state. 
For each case, both configurations 
C1 (filled circle) and C2 (open circle) 
are shown.
The number of copies $N$ is fixed at $10^5$
which is believed large enough. 
}
\label{fig3}
\end{figure}

In the presence of noise, 
the state we measure $|\psi'\rangle$,
is different from the initial state $|\psi\rangle$.
Therefore, as a consequence, 
$\overline{D}(\psi,\widetilde\psi')$ 
will first decrease 
and then quickly comes to saturate when increasing  $N$.
Furthermore, as can be seen from the figure, 
$\overline{D}(\psi,\widetilde\psi')$ 
in C2 have a larger differential 
(the difference between with and without noise) 
in comparison to C1 case. This remark implies that
C2 is more fragile against the noise than C1 because, in C2,
the state-postselection noise is presented in the interaction 
($|\mathfrak{c}_0'\rangle$ in Eq.~\eqref{eq:U-i}) 
while in C1, this noise is
mostly discarded through the postselection process. 
Interestingly, under the presence of noise, 
the systematic error is broken: 
there are no different in the measurement accuracies 
for the two configurations.

To further investigate the effect of noise,
we examine $\overline{D}(\psi,\widetilde\psi')$
 as a function of 
the noise parameter $\sigma$. 
Here, we consider the random and $\text{GHZ}_3$ cases 
as they are well established,  
for example, see \cite{Carvacho2017, Li:19}.
These other cases can be analyzed similarly.  
The results are shown in Fig.~\ref{fig3} 
for a fixed $N =10^5$.
Obviously, in the absence of noise, the result from C2 is 
better than that one from C1. 
Whenever the noise increases,
$\overline{D}(\psi,\widetilde\psi')$ from C2 
quickly increases and coincides with that one from C1
as an effect of the noise sensitivity as we discussed above.
As a result, the systematic error is broken out when 
increasing the noise parameter $\sigma$.



\subsection{Numerical results for mixed states}

For the mixed-state case, we consider the GHZ state 
is transformed under the 
white noise
such that
$\rho'_0 = (1-\epsilon)|\text{GHZ}_3\rangle\langle
\text{GHZ}_3|+\epsilon \bm{I}_s/8$,
where $0 \le \epsilon \le 1$ is the noise parameter. 
Such white noise transforms the state from 
the original (without noise)
$\epsilon = 0$ to 
the maximum noise $\epsilon = 1$.
The trace distance is defined by
\begin{align}\label{trm}
D(\rho_0,\widetilde\rho) = \dfrac{1}{2}{\rm Tr}
\bigl|\widetilde\rho-\rho_0\bigr|.
\end{align}

\begin{figure} [t]
\centering
\includegraphics[width=8.6cm]{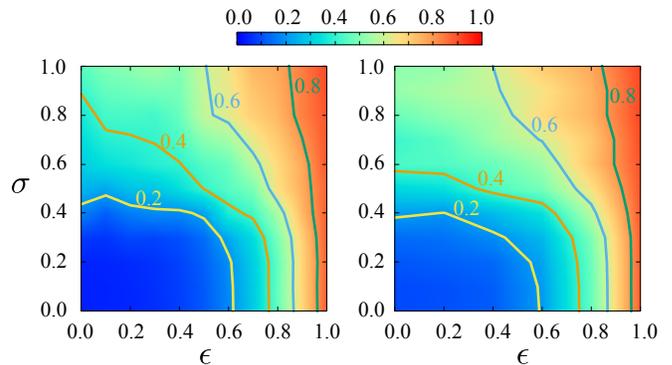}
\caption{
Density plot of the mean trace distance 
$\overline{D}(\rho_0,\widetilde\rho)$
as a function of $\epsilon$ and $\sigma$
for $N = 10^3$.
Two configurations C1 (left) 
and C2 (right)
are plotted and compared. 
}
\label{fig4}
\end{figure}

In Fig.~\ref{fig4},
we examine the mean trace distance
$\overline{D}(\rho_0,\widetilde\rho)$
over the noisy state-preparation parameter
$\epsilon$ and the 
noisy state-postselection parameter $\sigma$.
In genereal,
$\overline{D}(\rho_0,\widetilde\rho)$
increases when increasing 
$\epsilon$ and $\sigma$. 
Notably, when the state-preparation 
reaches the maximum noise 
$\epsilon = 1$, then 
$\overline{D}(\rho_0,\widetilde\rho)$
also reaches the maximum
regardless the noisy state-postselection
$\sigma$. 
Obviously, the noisy state-preparation
plays a decisive role in the accuracy 
of the tomography process. 

\begin{figure} [t]
\centering
\includegraphics[width=8.6cm]{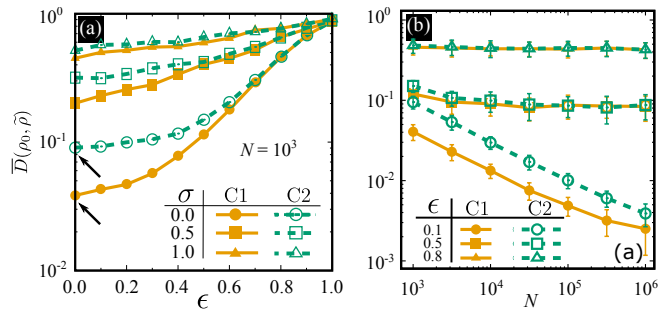}
\caption{
(a) Plot of 
$\overline{D}(\rho_0,\widetilde\rho)$
versus $\epsilon$ for several $\sigma$'s
as shown in the figure.
The results are extracted from Fig.~\ref{fig4}.
Two configurations C1 
and C2
are plotted and compared. 
The arrows indicate the case
$\epsilon = \sigma = 0$.
(b) Log-log plot of $\overline{D}(\rho_0,\widetilde\rho)$
as a function
of $N$ for $\epsilon = 0.1, 05,$ and 0.8 at fixed 
$\sigma = 0.05$.
}
\label{fig5}
\end{figure}

To look more detailed the accuracy in the two 
configurations C1 and C2, 
we extracted out some values from 
Fig.~\ref{fig4}
and plot them in Fig.~\ref{fig5}(a).
First, let us consider the noiseless case, 
i.e., $\epsilon = 0$
and $\sigma = 0$,
as indicated by the arrows 
in Fig.~\ref{fig5}(a).
In this case, 
$\overline{D}(\rho_0,\widetilde\rho)$
for C1 is better than C2.
Again, this is the systematic error
caused by different configurations. 

When the noises are presented
($\epsilon \neq 0$ and $\sigma \neq 0$), 
$\overline{D}(\rho_0,\widetilde\rho)$ 
in C1 quickly increases and reaches 
that increasing in C2. 
For $\epsilon \to 1$,
all $\overline{D}(\rho_0,\widetilde\rho)$s
converge regardless $\sigma$
as we have mentioned above. 
This observation implies that
when the SPAM errors appear
and large enough, 
the systematic error will be broken out.
For supporting our argument,  
we plot $\overline{D}(\rho_0,\widetilde\rho)$
as a function of $N$ in Fig. \ref{fig5}(b) 
for several $\epsilon$'s at
fixed $\sigma = 0.05$. 
It can be seen that $\overline{D}(\rho_0,\widetilde\rho)$ 
behaves similarly to  the pure-state case, 
where it comes to saturate (and converge) 
for large $N$ whenever the noise is presented. 
Obviously, when increasing $N$ large enough
to eliminate the statistical error, 
the systematic error
will be broken out by noises,
i.e., SPAM errors.

\section{Precision of 
quantum state tomography under noise}\label{seciv}
We analytically investigate the impact  
of the state-preparation error
on the precision of quantum state tomography
by using the features of quantum Fisher information.
Typically, the estimation of a parameter $\theta$
through an estimator will result in another value 
($\tilde\theta$)
which is, in general, different from $\theta$.
The variance between $\theta$ and $\tilde\theta$ 
is then given by $\Delta^2\theta = 
\mathbb{E}(\tilde\theta^2)-\mathbb{E}^2(\tilde\theta)
$ ($\mathbb{E}$ the average value,) 
and has a lower bound imposed by quantum mechanics:
\begin{align}
\Delta^2\theta \ge \dfrac{1}{Q_\theta},
\end{align}
is called quantum Cram\'er-Rao bound, where 
$Q_\theta$ is the quantum Fisher information (QFI)
corresponding to the maximization 
over all possible measurements
of the optimal estimator~\cite{Paris2004}.  
This is the ultimate achievable precision 
in the estimation of $\theta$.

In our pure-state case, 
for a given quantum state without noise
$|\psi\rangle = \sum_n \psi_n |n\rangle$.
Without loss of generality, 
we can assume $\psi_n$ is real
because the real part
and imaginary part of 
a wave function can be evaluated separately.
From the quantum estimation theory, 
for each $n$, we evaluate
the QFI by
\begin{align}\label{eq:qfi}
Q_n = 
\langle\psi|\bm L^2|\psi\rangle
= 4\Bigl[1-|\psi_n|^2\Bigr],
\end{align}
where $\bm L = 2(|\partial_{\psi_n}\psi\rangle
\langle\psi|+|\psi\rangle\langle\partial_{\psi_n}\psi|)$
is the symmetric logarithmic derivative
\cite{doi:10.1142/S0219749909004839}.
The total QFI is $Q = \sum_n Q_n = 4(d-1)$,
which is the maximum information 
that the measurements
can gain 
for the given state $|\psi\rangle$.

Similarly, for a quantum state contains some noises given by
$|\psi'\rangle = \dfrac{1}{\mathcal N}\sum_n
\bigl(\psi_n + \delta_n\bigr)|n\rangle$ as in Eq.~\eqref{e:pp},
the total QFI is (see App.~\ref{appD})
\begin{align}\label{eq:tqfi}
Q' = \dfrac{4}{\mathcal{N}^2}
(d-1),
\end{align}
where $\mathcal{N}$ is the normalization constant.
In the absence of noise, $\mathcal{N} = 1$
and thus, $Q' = Q$.

\begin{figure} [t]
\centering
\includegraphics[width=8.6cm]{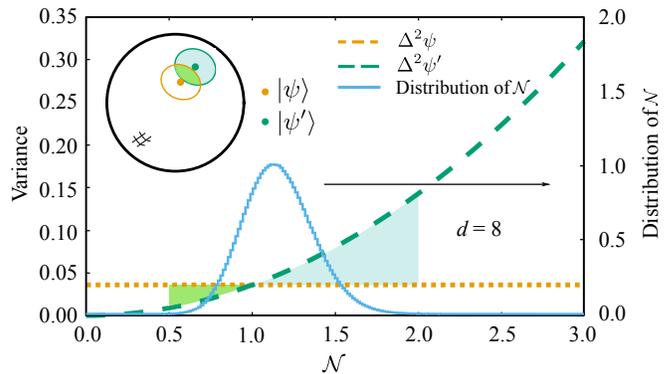}
\caption{
The variances $\Delta^2$ for two cases of without noise
($\Delta^2\psi$)
and with noise ($\Delta^2\psi'$)
as functions of the normalization  constant 
$\mathcal{N}$. Here,
$\mathcal{N}$ obeys
the normal distribution as 
numerically shown by the cyan normalized histogram
(bottom-right axes.) 
The histogram was obtained from a random state 
for $d = 8$.
Highlighted areas represent the differential between 
$\Delta^2\psi$ and $\Delta^2\psi'$ curves
and can be calculated by taking the integral
under these curves 
bounded by
a certain range of non-zero $\mathcal{N}$.
Inset: an illustration of 
distribution regions for the true state
and the state-preparation error. 
Ideally, the reconstructed states 
will locate around the measured state,
which forms a so-call distribution region. 
The orange dot and its rounded circle are 
the measured state and the distribution region
when the measured state is the true state $|\psi\rangle$,
likewise for the state-preparation error ($|\psi'\rangle$)
indicated by the green dot and the distribution region
is the green circle.
}
\label{fig6}
\end{figure}

To evaluate the efficiency of the measurement,
we can define the variances $\Delta^2\psi \propto Q^{-1}$
and $\Delta^2\psi' \propto [Q']^{-1}$ for cases of 
without and with noise, respectively. In Fig.~\ref{fig6},
we examine the variances versus $\mathcal{N}$
for $d = 8$.
It can be seen that $\Delta^2\psi = 1/28$ is a constant, 
while $\Delta^2\psi' < \Delta^2\psi$ for $\mathcal{N} < 1$
and $\Delta^2\psi' \ge \Delta^2\psi$ for $\mathcal{N} \ge 1$.
In other words, the state-preparation error 
can gain more information for $\mathcal{N} < 1$ 
while it loses information 
for $\mathcal{N} > 1$.

Furthermore, for a normal distribution of $\delta_n$ in
$|\psi'\rangle$,
the normalization  constant $\mathcal{N}$ obeys the  
normal distribution as we illustrate by the cyan curve 
in Fig.~\ref{fig6} for a random state $d = 8$.
Suppose we compare these results
for two cases of $\mathcal{N} < 1$ and $\mathcal{N} > 1$
under the statistical distribution of $\mathcal{N}$
(highlighted areas),  
it always shows $\Delta^2\psi' \ge \Delta^2\psi$.
The highlighted areas are given by 
the integral under the curves 
 $\Delta^2\psi$ and $\Delta^2\psi'$ and bounded by
a certain range of non-zero $\mathcal{N}$,
i.e., $\mathcal{N} \in [0.5, 2.0]$,
as can be seen from the figure.
This result implies that the variance with noise is 
larger than that one without noise. 
When increasing the noise parameter $\sigma$,
the distribution of $\mathcal{N}$ 
shifts toward the right,
which results in the increasing of 
$\Delta^2\psi' $ in the right highlighted area 
and decreasing in the left highlighted area, 
and thus losing precision.

In the inset figure, we illustrate the distribution regions
for these two cases of $|\psi\rangle$ and $|\psi'\rangle$. 
Here, the orange dot is the true state $|\psi\rangle$, 
and the orange circle is the distribution region 
where a reconstructed states 
$|\tilde\psi\rangle$ are located,
likewise for the green dot indicates 
the noisy state $|\psi'\rangle$
and the distribution region 
of the reconstructed states 
$|\tilde\psi'\rangle$ is the green circle.
There are situations (indicated by the lime area) 
that a reconstructed state
 $|\tilde\psi'\rangle$ is better than $|\tilde\psi\rangle$. 
These cases link to $\mathcal{N} < 1$ in the main figure.
Inversely, the cyan area indicates those situations that 
$|\tilde\psi'\rangle$ is worse than $|\tilde\psi\rangle$,
which connects to $\mathcal{N} > 1$ in the main figure. 
The illustration of these distribution regions thus 
directly leads to two different cases for $\mathcal{N}$,
as can be seen from the main figure.

Overall, the state-preparation error
gives less precision  
whenever the noise is presented 
regardless of how good   
measurement equipment is. 
We expect that appropriate optimal techniques, 
including quantum error correction 
and neural network architectures, 
could eliminate the SPAM errors
\cite{doi:10.1098/rspa.1998.0160,PhysRevLett.104.020401,Palmieri2020}.

\section{Conclusion}\label{secv}
We have numerically and 
analytically investigated the  
efficiency of the direct quantum state
measurement (DSM) under the
state-preparation-and-measurement (SPAM) errors
by using quantum-controlled measurements.
We found that when the noisy state-preparation
is presented, 
the measurement scheme gives less precise 
due to the bias between the true state and 
the state to be measured 
(the state that contains the noise.)
Nevertheless, the noisy state-postselection has
a significant effect on the measurement 
configurations, especially when 
combining with scan-free techniques in the DSM.
Furthermore, under such a SPAM error, 
the systematic error will be broken,
which results in the same accuracy for
both configurations.  
Our study could
provide an urgent outcome for
understanding the effect of SPAM errors
on quantum state tomography (QST).
In comparison to 
the conventional QST,
the DSM is more reliable 
in some particular cases such as high-dimension systems
\cite{Shi:15,Malik2014,doi:10.1002/lpor.201900251}
and nonlocal entangled 
tomography~\cite{PhysRevLett.123.150402}.
However, a detailed comparison of 
the DSM efficiency and the conventional QST
under noise requires much more investigation 
on quantum estimation theory. 
Thus, we leave it to the subject of future work.
Further studies in this field also 
include error correction schemes 
and neural network architectures 
to eliminate SPAM errors.

\begin{acknowledgments}
We would like to thank the anonymous referees 
for their valuable comments and suggestions.
This work was supported by 
JSPS KAKENHI Grant Number 20F20021 and 
the Vietnam National University under 
Grant Number QG.20.17.
LBH would like to thank Shikano for pointing out
Ref. \cite{Shikano_2009}.
\end{acknowledgments}

\appendix
\setcounter{equation}{0}
\renewcommand{\theequation}{A.\arabic{equation}}
\section{An example of noisy 
quantum state preparation}\label{appA}
\begin{figure} [t]
\centering
\includegraphics[width=8.6cm]{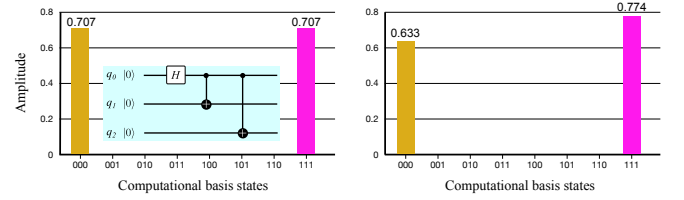}
\caption{
Inset: Quantum circuit for generating GHZ$_3$ state. 
Initially, three qubits
 $q_0, q_1$, and $q_2$
are prepared in the ground state, i.e., $|000\rangle$. 
Quantum Hadamard gate and 
Control-NOT gates are applied to
transform the initial state to the desired state. 
See detailed in Appendix~\ref{appA}.
Left: the amplitudes of the components in the GHZ$_3$ state 
after applying $H$ and $CNOT$ gates onto 
the initial state $|000\rangle$.
Right: the amplitudes of the components in the GHZ$_3$ state 
under the imperfection of the Hadamard gate.
These amplitudes are lightly deviated 
from the true values in the left figure.
}
\label{fig7}
\end{figure}

We provide an example for 
preparing the quantum state GHZ$_3$ that contains noise.
Consider a quantum circuit as shown in
Fig.~\ref{fig7} (inset).
Therein, three qubits $q_0, q_1$,
and $q_2$ are prepared 
in the ground state, i.e., $|000\rangle$.
Applying a sequence of Hadamard ($H$) gate onto $q_0$,
control-NOT ($CNOT$) gate onto $q_0, q_1$, 
and control-NOT gate onto $q_0, q_2$, 
as shown in the inset figure, respectively,
we obtain the output state as
\begin{align}\label{eq:ghz}
\notag |000\rangle \xrightarrow[]{H} 
&\dfrac{1}{\sqrt{2}}(|000\rangle+|100\rangle)
\xrightarrow[]{CNOT}  \dfrac{1}{\sqrt{2}}(|000\rangle+|110\rangle)\\
&\xrightarrow[]{CNOT}  \dfrac{1}{\sqrt{2}}(|000\rangle+|111\rangle),
\end{align}
which is the GHZ$_3$ state.
We simulate this state in 
Fig.~\ref{fig7} (left) by using the IBM Qiskis package. 
It can be seen that the amplitudes of 
$|000\rangle$ and $|111\rangle$ are the same and 
equal to $1/\sqrt{2}$.

Now, let us assume the imperfection 
in the Hadamard gate as follows.
We first decompose the Hadamard gate
into the two rotations:
$\pi/2$ about the Y-axis, and  
$\pi$ about the Z-axis, such that
\begin{align}\label{ha}
-iH = 
\begin{pmatrix}
\cos(\pi/4) & -\sin(\pi/4) \\
\sin(\pi/4)&\cos(\pi/4)
\end{pmatrix}
\begin{pmatrix}
e^{-i\pi/2} & 0 \\
0 & e^{i\pi/2}
\end{pmatrix},
\end{align}
where the rotation matrices are
given as
\begin{align}\label{matrics}
R_y(\theta) &= 
\begin{pmatrix}
\cos(\theta/2) & -\sin(\theta/2) \\
\sin(\theta/2)&\cos(\theta/2)
\end{pmatrix} \;, \\ 
R_z(\theta) &= 
\begin{pmatrix}
e^{-i\theta/2} & 0 \\
0 & e^{i\theta/2}
\end{pmatrix}.
\end{align}

Under the imperfection, 
assume that the rotation angles will deviate 
from their true values, 
such as $\pi/2 + \alpha$,
and $\pi + \beta$,
where $\alpha$ and $\beta$ are small angles.
Without loss of generality,
we can choose $\beta = 0$ since 
the operation of $R_z(\theta)$ does not 
affect the amplitudes of the quantum state. 
As a result, the Hadamard gate becomes
\begin{align}\label{eq:hp}
\notag-iH' &= \dfrac{1}{\sqrt{2}}
\begin{pmatrix}
\cos(\frac{\alpha}{2})-\sin(\frac{\alpha}{2})
& 
\cos(\frac{\alpha}{2})+\sin(\frac{\alpha}{2})\\
\cos(\frac{\alpha}{2})+\sin(\frac{\alpha}{2})&
-[\cos(\frac{\alpha}{2})-\sin(\frac{\alpha}{2})]
\end{pmatrix} \\
&= \dfrac{1}{\sqrt{2}}
\begin{pmatrix}
a & b \\
b & -a 
\end{pmatrix}.
\end{align}
Therefore, the final state becomes
\begin{align}\label{eq:ghzp}
\dfrac{1}{\sqrt{2}}(a|000\rangle+b|111\rangle),
\end{align}
which slightly different 
from the true GHZ$_3$ state. Here, 
$\frac{1}{2}(|a|^2 + |b|^2) = 1$.
This is an example of a noisy state-preparation process. 
In Fig.~\ref{fig7} (right), we simulate the GHZ$_3$
state assuming that the Hadamard gate is under the imperfection 
at $\alpha = 0.2$ as an example. 
Under this noisy state-preparation, the two components 
$|000\rangle$ and $|111\rangle$ are deviated from $1/\sqrt{2}$.
In general, all the components will be deviated from their values
as we have modeled from Eq.~\ref{e:pp} in the main text.


\setcounter{equation}{0}
\renewcommand{\theequation}{B.\arabic{equation}}
\section{Quantum controlled measurements 
for pure states}\label{appB}
In this appendix, we closely follow 
the quantum controlled measurement framework 
introduced by Ogawa et al. \cite{Ogawa_2019} 
for pure states,
and we derive it under the noise 
using our denotation to 
make the work self-consistency.

We consider the initial joint state 
of the target system and 
the control qubit probe as 
$|\Psi\rangle = |\psi'\rangle\otimes|\xi\rangle$, where
\begin{align}
&|\psi'\rangle = \dfrac{1}{\mathcal N}\sum_{n=0}^{d-1}
\Bigl(\psi_n + \delta_n\Bigr)|n\rangle
= \sum_{n=0}^{d-1} \psi'_n |n\rangle 
, \text{ and } \label{app:psi}\\
&|\xi\rangle \equiv |+\rangle = \dfrac{1}{\sqrt{2}}
\Bigl(|0\rangle + |1\rangle\Bigr), \label{app:xi}
\end{align}
where $\mathcal{N}$ is the normalization factor, $\psi'_n = \frac{\psi_n+\delta_n}{\mathcal{N}}$,
and $\delta_n = x_1 + ix_2$ 
is a complex random number, 
where $x_1$ and $x_2$ are random numbers
that follow the normal distrbution 
$f(x) = \frac{1}{\sigma\sqrt{2\pi}}
\exp[-\frac{1}{2}(\frac{x}{\sigma})^2]$.

\subsection*{For C1}
For C1, following Ogawa et al. \cite{Ogawa_2019}, 
we consider the interaction as
\begin{align}\label{app:U-i}
\bm U_n = 
\bigl(\bm{I}_{\rm s}-|n\rangle\langle n|\bigr)
\otimes |0\rangle\langle 0| + |n\rangle\langle n|
\otimes |1\rangle\langle 1|.
\end{align}
After the interaction, the joint state becomes
\begin{align}\label{app:Psi'}
\bm U_n|\Psi\rangle=
\dfrac{1}{\sqrt{2}} \Bigl[
\sum_{m=0}^{d-1}\psi'_{m}|m\rangle- \psi'_n|n\rangle\Bigr]\otimes |0\rangle
+\dfrac{1}{\sqrt{2}} \psi'_n|n\rangle\otimes|1\rangle.
\end{align}
We postselect the target system onto the conjugate basis
\begin{align}\label{app:co}
|\mathfrak{c}'_0\rangle = \frac{1}{\mathcal{M}}\sum_{m=0}^{d-1}(1+\kappa_m)|m\rangle
= \sum_{m=0}^{d-1} \mathfrak{c}_m |m\rangle,
\end{align}
where $\mathcal{M}$ is the normalization factor, and $\mathfrak{c}_m = \frac{1+\kappa_m}{\mathcal{M}}$.
Here, $\kappa_m$ is a real random number 
and distributes according to the normal distribution.
The final control qubit state is given by
\begin{align}\label{eq:final}
\notag |\eta\rangle &= 
\bigl(\langle \mathfrak{c}'_0|
\otimes \bm {I}_{\rm p}\bigr)
\bm U_n|\Psi\rangle \\ 
&=\dfrac{1}{\sqrt{2}}\Bigl[
\big(\Gamma - \mathfrak{c}_n\psi'_n\big)|0\rangle
+\mathfrak{c}_n\psi'_n|1\rangle\Bigr],
\end{align} 
where $\Gamma = \sum_{m=0}^{d-1}
\mathfrak{c}_{m}\psi'_{m}$.

Finally, we measure the control qubit probe 
in the Pauli basis
$|j\rangle \in \{ |0\rangle, |1\rangle, 
|+\rangle, |-\rangle, |L\rangle, |R\rangle\}$,
where 
$|\pm\rangle = \frac{1}{\sqrt{2}}\bigl(|0\rangle\pm|1\rangle\bigr),
|L\rangle = \frac{1}{\sqrt{2}}\bigl(|0\rangle + i|1\rangle\bigr), 
|R\rangle = \frac{1}{\sqrt{2}}\bigl(|0\rangle - i|1\rangle\bigr)$.
The probability for measuring $|j\rangle\langle j|$ is $P_j = |\langle j|\eta\rangle|^2$ explicitly give
\begin{align}\label{eq:P}
	&P_0=\frac{1}{2}
	\Bigl[\Gamma^2 - 2\mathfrak{c}_n\Gamma \, 
	{\rm Re}\psi'_n 
	+\mathfrak{c}_n^2|\psi'_n|^2\Bigr],\\
    &P_1=\frac{1}{2}\mathfrak{c}_n^2|\psi'_n|^2,\\
	&P_+=\frac{1}{4}\Gamma^2,\\
	&P_-=\frac{1}{4}
	\Bigl[\Gamma^2-4\Gamma \mathfrak{c}_n \, {\rm Re}\psi'_n
	+4 \mathfrak{c}_n^2|\psi'_n|^2
	\Bigr],\\
	&P_L=\frac{1}{4}
	\Bigl[\Gamma^2-2\Gamma \mathfrak{c}_n \, 
	{\rm Re}\psi'_n
	+2\Gamma \mathfrak{c}_n \, 
	{\rm Im}\psi'_n + 2\mathfrak{c}_n^2|\psi'_n|^2
	\Bigr],\\
	&P_R=\frac{1}{4}
	\Bigl[\Gamma^2-2\Gamma \mathfrak{c}_n \, 
	{\rm Re}\psi'_n
	-2\Gamma \mathfrak{c}_n \, 
	{\rm Im}\psi'_n + 2 \mathfrak{c}_n^2 |\psi'_n|^2
	\Bigr].
\end{align}
As a result, the real and imaginary parts of the amplitude 
$\psi'_n$ are reproduced as
\begin{align}
{\rm Re}\psi'_n=\frac{P_+-P_-+2P_1}
{\mathfrak{c}_n\Gamma}, \ 
{\rm Im}\psi'_n=\frac{P_L-P_R}
{\mathfrak{c}_n\Gamma}.
\end{align}

\subsection*{For C2}
For C2, the interaction is given by \cite{Ogawa_2019}
\begin{align}\label{app:U-i}
\bm U=(\bm{I}_s-|\mathfrak{c}'_0\rangle
\langle \mathfrak{c}'_0|)\otimes|0\rangle\langle0|
+|\mathfrak{c}'_0\rangle
\langle \mathfrak{c}'_0|\otimes|1\rangle\langle1|.
\end{align}
The joint state after the interaction is given by
\begin{align}\label{app:final-j}
\notag \bm U|\Psi\rangle &=
\frac{1}{\sqrt{2}}\Bigl[\sum_{m=0}^{d-1}\psi'_{m}|m\rangle
    -\sum_{m=0}^{d-1}\psi'_{m}
    \mathfrak{c}_{m}|\mathfrak{c}'_0\rangle\Bigr]
    \otimes |0\rangle \\
&+\frac{1}{\sqrt{2}}\Bigl[\sum_{m=0}^{d-1}\psi'_{m}
    \mathfrak{c}_{m}|\mathfrak{c}'_0\rangle\Bigr]
\otimes |1\rangle.
\end{align}
After the interaction, the target system is postselected 
onto $|n\rangle$ while 
the remaining state of the control qubit probe
is given as
\begin{align}\label{eq:f-II}
    |\eta\rangle=\dfrac{1}{\sqrt{2}}\Bigl[\big(
    \psi'_n-\mathfrak{c}_n\Gamma\big)|0\rangle+
    \mathfrak{c}_n\Gamma|1\rangle\Big].
\end{align}
Measuring the control qubit probe in the Pauli basis
as above, we obtain
\begin{align}
P_0 &=
\dfrac{1}{2}\Bigl[|\psi'_n|^2-
2\mathfrak{c}_n\Gamma{\rm Re}\psi'_n
+\mathfrak{c}_n^2\Gamma^2\Bigr],\\
P_1&=
\dfrac{1}{2}\mathfrak{c}_n^2\Gamma^2,\\
P_+&=\frac{1}{4}|\psi'_n|^2,\\
P_-&=
\dfrac{1}{4}\Bigl[|\psi'_n|^2-4
\mathfrak{c}_n\Gamma{\rm Re}\psi'_n
+4 \mathfrak{c}_n^2\Gamma^2\Bigr],\\
P_L&=
\dfrac{1}{4}\Bigl[|\psi'_n|^2-2
\mathfrak{c}_n\Gamma{\rm Re}\psi'_n
+2\mathfrak{c}_n\Gamma{\rm Im}\psi'_n
+2\mathfrak{c}_n^2\Gamma^2\Bigr],\\
P_R&=
\dfrac{1}{4}\Bigl[|\psi'_n|^2-
2\mathfrak{c}_n\Gamma{\rm Re}\psi'_n
-2\mathfrak{c}_n\Gamma{\rm Im}\psi'_n
+2\mathfrak{c}_n^2\Gamma^2\Bigr].
\end{align}
Then, we have
\begin{align}
{\rm Re}\psi'_n=\frac{P_+-P_-+2P_1}
{\mathfrak{c}_n\Gamma},\
{\rm Im}\psi'_n=\frac{P_L-P_R}
{\mathfrak{c}_n\Gamma}.
\end{align}

\setcounter{equation}{0}
\renewcommand{\theequation}{C.\arabic{equation}}
\section{Quantum controlled measurements for mixed states}
\label{appC}

We consider the joint state $\Lambda$ as following
\begin{align}
    \Lambda = \rho'_0 \otimes |+\rangle\langle +|, \quad 
    \text{with}\; \rho'_0 = \sum_{n,m=0}^{d-1}\rho'_{nm}|n\rangle\langle m|.
\end{align}

\subsection*{For C1}
The interaction operator is given the same as above:
\begin{align}\label{app:ur}
\bm U_n = \big(\bm{I}_{\rm s} - |n\rangle\langle n|\big) \otimes |0\rangle\langle 0 |+ |n\rangle\langle n|\otimes |1\rangle\langle 1|.
\end{align}
After the interaction, the joint state evolves to
\begin{align}
    \Lambda' = \bm U_n\Lambda \bm U_n^{\dag}\, ,
\end{align}
which is explicitly written as
\begin{align}
\notag \Lambda' &= \Big[\rho'_0 -\Bigl(\sum_{m=0}^{d-1}\rho'_{nm}|n\rangle\langle m|+c.c\Bigr)+\rho'_{nn}|n\rangle\langle n|\Big]\otimes \frac{1}{2}|0\rangle\langle 0|\\
& + \Big[\sum_{m=0}^{d-1}\rho'_{mn}|m\rangle\langle n|-\rho'_{nn}|n\rangle\langle n|\Big]\otimes \frac{1}{2}|0\rangle\langle 1|\notag\\
& + \Big[\sum_{m=0}^{d-1}\rho'_{nm}|n\rangle\langle m|-\rho'_{nn}|n\rangle \langle n|\Big]\otimes \frac{1}{2}|1\rangle\langle 0|\notag\\
& + \Big[\rho'_{nn}|n\rangle\langle n|\Big]\otimes\frac{1}{2}|1\rangle\langle 1|.
\end{align}
Here, $c.c$ stands for `complex conjugate.'
After postselecting this state onto 
\begin{align}\label{app:k_c1_rho}
\notag|\mathfrak{c}'_k\rangle\langle \mathfrak{c}'_k| &= 
\frac{1}{\mathcal{M}^2}\sum_{n,m=0}^{d-1} e^{i2\pi (m-n)k/d}
(1+\kappa_{m})(1+\kappa_{n})|m\rangle\langle n|\\
&=\sum_{n,m = 0}^{d-1} e^{i2\pi (m-n)k/d}
\mathfrak{c}_m\mathfrak{c}_n
|m\rangle\langle n|\;,
\end{align}
the final state
of the control qubit probe becomes
\begin{align}\label{app:fp}
    \Lambda'' = \langle \mathfrak{c}'_k|
    \Lambda'|\mathfrak{c}'_k\rangle = 
    \begin{pmatrix}
        \Lambda_{00}''(n,k) & \Lambda_{01}''(n,k) \\
        \Lambda_{10}''(n,k) & \Lambda_{11}''(n,k)
    \end{pmatrix}.
\end{align}
Explicitly,
\begin{align}
\notag\Lambda_{00}''(n,k) &= \frac{1}{2}\Big[\sum_{n,m=0}^{d-1}
\rho'_{nm}e^{i2\pi (m-n)k/d} \mathfrak{c}_{m} \mathfrak{c}_{n} \\
&\hspace{-1.5cm}-\Big(\sum_{m=0}^{d-1}
\rho'_{nm}e^{i2\pi (m-n)k/d}\mathfrak{c}_{n} 
\mathfrak{c}_{m}+c.c\Big)
+\rho'_{nn} \mathfrak{c}_n^2 \Big] \label{app:rho00}\\
\Lambda_{01}''(n,k) &= \frac{1}{2}\Big[
\sum_{m=0}^{d-1}\rho'_{mn}e^{i2\pi (n-m)k/d}
\mathfrak{c}_n \mathfrak{c}_{m}-\rho'_{nn} 
\mathfrak{c}_n^2 \Big] \label{app:rho01}\\
\Lambda_{10}''(n,k) &= [\Lambda_{01}''(n,k)]^* \label{app:rho10}\\
\Lambda_{11}''(n,k) &= \frac{1}{2}\rho'_{nn}
\mathfrak{c}_n^2. \label{app:rho10}
\end{align}

Using Fourier transformation on $\Lambda''_{10}(n,k)$, 
we obtain 
\begin{align}\label{app:rc}
\rho'_{nm} \varpropto \frac{1}{\mathfrak{c}_{n}\mathfrak{c}_{m}}
\Big[d\delta_{n,m}
\Lambda''_{11}(n,k)+\sum_{k=0}^{d-1}e^{i2\pi (n-m)k/d}
\Lambda_{10}''(n,k) \Big].
\end{align}
To get $\Lambda_{10}''(n,k)$ and $\Lambda''_{11}(n,k)$, the control
qubit is measured as follows:
\begin{align}
    &\Lambda''_{10}(n,k) = \frac{1}{2}\Big[(P_+ - P_-) + i(P_L-P_R) \Big],
    \, \text{and}\\
    &\Lambda''_{11}(n,k) = P_1,
\end{align}
where $P_j = \Tr [|j\rangle\langle j|\Lambda'']$
is the probability when measuring the control qubit probe
in the element $j$ of the Pauli basis.

\subsection*{For C2}
In this case, the interaction is
\begin{align}\label{app:uc2} 
\bm U = \big({\bm I}_{\rm s} - |\mathfrak{c}'_k\rangle
\langle \mathfrak{c}'_k|\big) \otimes 
|0\rangle\langle 0 |+ |\mathfrak{c}'_k\rangle
\langle \mathfrak{c}'_k|\otimes
|1\rangle\langle 1|.
\end{align}
After applying this interaction $U$, the initial joint state
becomes
\begin{align}
\Lambda' = \bm U\Lambda \bm U^{\dag},
\end{align}
which is explicitly given as
\begin{widetext}
\begin{align}
\notag \Lambda' &= \Big[\rho'_{0} - \sum_{n,m=0}^{d-1}
    \rho'_{nm}\mathfrak{c}_{m} e^{i2\pi km/d}|n\rangle 
    \langle \mathfrak{c}'_k | 
    - |\mathfrak{c}'_k\rangle \sum_{n,m=0}^{d-1}
    \rho'_{nm}\mathfrak{c}_{n} e^{-i2\pi kn/d}\langle m| \\
\notag    &\hspace{5cm}
    + \sum_{n,m=0}^{d-1}
    \rho'_{nm}\mathfrak{c}_{n}\mathfrak{c}_{m} e^{i2\pi k(m-n)/d}
    |\mathfrak{c}'_k\rangle \langle \mathfrak{c}'_k|
    \Big]\otimes \frac{1}{2}|0\rangle\langle 0|\\
    & + \Big[ \sum_{n,m=0}^{d-1}
    \rho'_{nm}\mathfrak{c}_{m} e^{i2\pi km/d}|n\rangle 
    \langle \mathfrak{c}'_k|
     - \sum_{n,m=0}^{d-1}
    \rho'_{nm}\mathfrak{c}_{n}\mathfrak{c}_{m} e^{i2\pi k(m-n)/d}
    |\mathfrak{c}'_k\rangle \langle \mathfrak{c}'_k|\notag
    \Big] \otimes \frac{1}{2}|0\rangle\langle 1|\notag\\
    & + \Big[|\mathfrak{c}'_k\rangle \sum_{n,m=0}^{d-1}
    \rho'_{nm}\mathfrak{c}_{n} e^{-i2\pi kn/d}\langle m| \notag
    - \sum_{n,m=0}^{d-1}
    \rho'_{nm}\mathfrak{c}_{n}\mathfrak{c}_{m} e^{i2\pi k(m-n)/d}
    |\mathfrak{c}'_k\rangle \langle \mathfrak{c}'_k|\notag
        \Big]\otimes \frac{1}{2}|1\rangle\langle 0|\notag\\
    & + \Big[\sum_{n,m=0}^{d-1}
    \rho'_{nm}\mathfrak{c}_{n}\mathfrak{c}_{m} e^{i2\pi k(m-n)/d}
    |\mathfrak{c}'_k\rangle \langle \mathfrak{c}'_k|
        \Big]\otimes\frac{1}{2}|1\rangle\langle 1|.
\end{align}
\end{widetext}
Next, we postselect this state onto $|n\rangle\langle n|$
and get
\begin{align}
    \Lambda'' = \langle n|\Lambda'|n\rangle = 
    \begin{pmatrix}
        \Lambda_{00}''(n,k) & \Lambda_{01}''(n,k) \\
        \Lambda_{10}''(n,k)& \Lambda_{11}''(n,k)
    \end{pmatrix}.
\end{align}
Explicitly,
\begin{align}
\notag\Lambda_{00}''(n,k) &= \frac{1}{2}\Big[
\rho'_{nn}
- \Big(\sum_{m=0}^{d-1}
\rho'_{nm}\mathfrak{c}_{m}\mathfrak{c}_{n}
e^{i2\pi k(m-n)/d} + c.c.\Big) \\
& + \Big(\sum_{n,m=0}^{d-1}
\rho'_{nm}\mathfrak{c}_{n}\mathfrak{c}_{m}
e^{i2\pi k(m-n)/d}
\Big)\mathfrak{c}_{n}^{2}\Big], \label{app:r00} \\
\notag \Lambda_{01}''(n,k) &=\frac{1}{2}\Big[
\Big(\sum_{m=0}^{d-1}
\rho'_{nm}\mathfrak{c}_{m}\mathfrak{c}_{n}
e^{i2\pi k(m-n)/d} \Big) \\
&- \Big(\sum_{n,m=0}^{d-1}
\rho'_{nm}\mathfrak{c}_{n}\mathfrak{c}_{m}e^{i2\pi k(m-n)/d}
\Big)\mathfrak{c}_{n}^{2}\Big], \label{app:r01} \\
\Lambda_{10}''(n,k) &= [\Lambda_{01}''(n,k)]^*, \label{app:r10} \\
\Lambda_{11}''(n,k) &= \frac{1}{2}
\Big[\sum_{n,m=0}^{d-1}
\rho'_{nm}\mathfrak{c}_{n}\mathfrak{c}_{m}e^{i2\pi k(m-n)/d}
\Big]\mathfrak{c}_{n}^{2}. \label{app:r11}
\end{align}
Using Fourier transformation on $\Lambda_{01}''(n,k)$, 
we obtain: 
\begin{align}\label{app:rc2}
\rho'_{nm} \varpropto \frac{1}{\mathfrak{c}_{m}\mathfrak{c}_{n}}
\Big[ \sum_{k=0}^{d-1} e^{i2\pi k(n-m)/d}
\Big(\Lambda''_{01}(n,k) + \Lambda''_{11}(n,k) \Big)
\Big],
\end{align}
where $\Lambda_{01}''(n,k)$ is obtained by measuring
the control qubit probe as follows:
\begin{align}
    \Lambda''_{01}(n,k) = \frac{1}{2}\Big[(P_+ - P_-) - i(P_L-P_R) \Big].
\end{align}

\setcounter{equation}{0}
\renewcommand{\theequation}{D.\arabic{equation}}
\section{Quantum Fisher Information}\label{appD}
In this section, we show how to calculate 
the total quantum Fisher information (QFI) 
for $|\psi'\rangle$ state:
\begin{align}\label{app:psi'}
|\psi'\rangle = \dfrac{1}{\mathcal{N}}
\sum_{n=0}^{d-1} \bigl(\psi_n + \delta_n\bigr) |n\rangle,
\end{align}
where $\psi_n$ is unknown.
The normalization  constant is
\begin{align}\label{app:ns}
\mathcal{N}^2 = \sum_{n=0}^{d-1}
\Bigl(\psi_n + \delta_n\Bigr)^2.
\end{align}
Here, note that we consider both 
$\psi_n$ and  $\delta_n$ are real
for simplicity.
First, we calculate $\partial_{\psi_n}|\psi'\rangle$,
where we are using $\partial_{\psi_n}$
as shorthand for $\partial/\partial{\psi_n}$.
We have
\begin{align}\label{app:ns}
\notag \partial_{\psi_n}|\psi'\rangle
& = \dfrac{\partial}{\partial{\psi_n}}
\Bigl(\dfrac{1}{\mathcal N}\Bigr)
\sum_{n=0}^{d-1} 
\bigl(\psi_n + \delta_n\bigr) |n\rangle \\
\notag &\hspace{2cm} + \dfrac{1}{\mathcal N}
\dfrac{\partial}{\partial{\psi_n}}
\Bigl(\sum_{n=0}^{d-1} 
\bigl(\psi_n + \delta_n\bigr) |n\rangle
\Bigr)
\\
& = -\dfrac{\psi_n+\delta_n}
{\mathcal N^2} |\psi'\rangle + 
\dfrac{1}{\mathcal N}
 |n\rangle.
\end{align}
The Quantum Fisher Information (QFI) is given by
\begin{align}\label{app:qfi'}
\notag Q'_n &= 4
\Bigl[\dfrac{\langle\partial\psi'|}{\partial _{\psi_n}}
\dfrac{|\partial\psi'\rangle}{\partial _{\psi_n}}
-\Bigl|\dfrac{\langle\partial\psi'|}
{\partial _{\psi_n}}|\psi'\rangle\Bigr|^2\Bigr],\\
&= 
\dfrac{4}{\mathcal N^2}
\Bigl[ 1 - 
\dfrac{(\psi_n+\delta_n)^2}
{\mathcal N^2}
\Bigr].
\end{align}

Then, the total QFI is
\begin{align}\label{app:t}
Q' = \sum_{n=0}^{d-1}Q'_n = 
\dfrac{4}{\mathcal N^2} (d-1).
\end{align}

\bibliography{mybib}

\end{document}